# Relational Blockworld: Towards a Discrete Graph Theoretic Foundation of Quantum Mechanics

*W.M. Stuckey[1], Timothy McDevitt[2] and Michael Silberstein[3,4]*

## Abstract

We propose a discrete path integral formalism over graphs fundamental to quantum mechanics (QM) based on our interpretation of QM called Relational Blockworld (RBW). In our approach, the transition amplitude is not viewed as a sum over all field configurations, but is a mathematical machine for measuring the symmetry of the discrete differential operator and source vector of the discrete action. Therefore, we restrict the path integral to the row space of the discrete differential operator, which also contains the discrete source vector, in order to avoid singularities. In this fashion we obtain the two-source transition amplitude over a "ladder" graph with *N* vertices. We interpret this solution in the context of the twin-slit experiment.

[1] Department of Physics, Elizabethtown College, Elizabethtown, PA 17022, stuckeym@etown.edu
[2] Department of Mathematics, Elizabethtown College, Elizabethtown, PA 17022, mcdevittt@etown.edu
[3] Department of Philosophy, University of Maryland, College Park, MD 20742
[4] Department of Philosophy, Elizabethtown College, Elizabethtown, PA 17022, silbermd@etown.edu

## 1. INTRODUCTION

It is now widely appreciated that the notion of causality as characterized by the light-cone structure of special relativity (SR) might be jeopardized by the combination of the relativity of simultaneity (RoS) and quantum non-locality (QNL), i.e, correlated, space-like separated experimental outcomes that violate Bell's inequality. The problem is of course that RoS does not allow for a definite temporal ordering of space-like separated events, yet correlated experimental outcomes in quantum mechanics (QM) violating Bell's inequality seem to suggest something like causal connections between outcomes, at least on some interpretations. If one accepts unambiguous temporal ordering as a necessary condition for causation, then apparently the combination of RoS and QNL threatens causation *a la* SR. While some might have us abandon RoS, there is an interpretative set of QM dubbed "time-symmetric approaches" that addresses this conflict head on. There are several variations on this theme (see the focus issue in *Studies in History and Philosophy of Modern Physics* **39**, Nov 08), but the basic idea is to employ a future boundary condition in spacetime, e.g., the actual outcomes of quantum experiments as in the path integral formalism.

In backwards-causation time-symmetric (BCTS) approaches, one eliminates the directional nature of a causal relationship so that there is no distinction between "A causes B" and "B causes A," but rather it is merely the case that "A and B are causally related." In this sense, the outcomes of QNL experiments are "causally related" to the state preparation so the demand for a causal relationship (per the violation of Bell's inequality) between the space-like separated, correlated outcomes is achieved by allowing for the fact that outcomes "influence" the state preparation (thus, the term "backwardly causal" although "bi-causal" might be more appropriate). BCTS provides for a local account of entanglement (one without space-like influences) that not only keeps RoS, but in some cases relies on it by employing its blockworld consequence—the reality of all events past, present and future including the outcomes of quantum experiments (Peterson & Silberstein, 2009; Silberstein *et al.*, 2007). Given the future boundary condition in spacetime, one is free to view configuration space (the wave function) as a mere calculational device (because we need only take the actual outcomes of experiments seriously), thus rendering the quantum and spacetime pictures fully harmonious[1].

We have shown (Silberstein *et al.*, 2008) that BCTS is not sufficient to account for all QNL experiments locally, e.g., the quantum liar experiment, and we have instead championed the *Relational Blockworld* (RBW) interpretation of QM. RBW is a completely acausal and adynamical interpretation providing a purely geometrical account of QNL that violates separability but not locality (Stuckey *et al.*, 2008). RBW takes seriously the blockworld perspective and further holds that QM systems are not composed fundamentally of *dynamical entities*, such as particles or waves evolving in time, but rather of relations. The name Relational Blockworld was coined since relations rather than relata are the fundamental "constituents" in a blockworld setting. As an acausal account, RBW rejects any kind of common-cause principle, i.e., the claim that every systematic quantum correlation between events is due to a cause that they share whether in the past or future. QM detector clicks are not evidence of microscopic dynamical entities (with ''thusness'' as Einstein would say) propagating through space and impinging on the detector. Rather, detector clicks evidence rarefied subsets of relations comprising the source, detector, beam splitters, mirrors, etc. in the entire worldtube of the experimental arrangement from initiation to outcomes (as in the case of entanglement), i.e., in an ''all at once'' (blockworld) fashion. Therefore, causality, dynamical entities and dynamical laws are emergent features in our view, not fundamental. In this way, we have been able to provide an account of QM that resolves all the foundational issues therein (Silberstein *et al.*, 2007 & 2008; Stuckey *et al.*, 2008).

Rather than rejecting RoS, we believe QM and SR together are telling us that what needs to be rejected is the dynamical picture as fundamental. That is, the most fundamental level of reality is not to be described via some fundamental entity or entities evolving in time according to dynamical laws against a spacetime background per certain boundary conditions. We are therefore led to conclude, as does Smolin (2006), that a new theory of physics is required to address the foundational problems of QM[2]. Of course there are many people looking for a theory fundamental to QM, but most are doing so with the explicit aim of unifying the quantum with general relativity—so called quantum gravity[3] (QG). In the arena of QG it is not unusual to find theories that are in some way underneath spacetime theory and theories of "matter" involving dynamical entities. However, the adynamical, acausal and relational nature of our interpretation of QM forces us to hunt for such a theory *just to underwrite QM itself*. Again, given the blockworld and relationalism of RBW, our fundamental account cannot be dynamical thus RBW contains a promissory note for a relational formalism underlying QM, in our case, one that is based on self consistency[4], not dynamical law (as explained in section 2 below). The formalism we have proposed, a path integral formalism over graphs, contains a discrete action constructed per a self-consistency criterion for dynamical entities, space and time. In a sense, we model QM as a spatially discrete quantum field theory (QFT), although we have in mind that a *spatiotemporally* discrete formalism, such as that presented, underlies *both* QM and QFT. Along those lines, our approach constitutes a *new basis* for QM as opposed to a mere *discrete approximation* thereto, since we are proposing an origin (self-consistency criterion) for the kernel Σ of the discrete action, which is otherwise fundamental. Previously, we presented a solution for the spatially discrete two-source transition amplitude after integration by parts (Stuckey *et al*., 2008). Herein we offer a spatiotemporally discrete solution for the two-source amplitude over a "ladder" graph of arbitrary size without assuming the boundary terms are well-behaved; in fact, we find the discrete differential operator of the action $\vec{\vec{A}}$ contains a zero eigenvalue.

[1] This view is in contrast to those giving ontic priority to configuration space, e.g., Many Worlds or Wheeler-DeWitt.

[2] Smolin writes (2006, 10), "The problem of quantum mechanics is unlikely to be solved in isolation; instead, the solution will probably emerge as we make progress on the greater effort to unify physics."

Given that our source vector $\vec{J}$ is orthogonal to the eigenvector corresponding to eigenvalue zero, i.e., resides in the row space of $\vec{\vec{A}}$, we deal with the singular nature of our transition amplitude $Z$ by restricting the integral to the ($N$-1)-dimensional row space of $\vec{\vec{A}}$ (as explained in section 3 below). This is justified by the fact that, per the blockworld view, $Z$ is not viewed as a "sum over all field configurations" in our discrete, acausal approach, but instead $Z$ is a 'mathematical machine' that measures the symmetry contained in $\vec{\vec{A}}$ and $\vec{J}$ (kernel) of the discrete action, as will be explained in section 2.

---

[3] Many regard relativistic quantum field theory (RQFT) as fundamental to both QM and SR, and hold that the advancement of physics requires a quantum 'version' of general relativity, i.e., quantum gravity (QG). Since RQFT does not, and QG need not, address the foundational problems of QM, our approach is unique in that we are starting with an interpretation of QM and working our way toward a theory of QG, motivated by the foundational problems of QM. While we do not explicitly address QG in this paper, one can readily see how our approach would change the nature of the QG program. At any rate, normally, the sort of discrete graph theoretic foundation we are alleging here for QM (fundamental even to spacetime and dynamical entities) would only be seen (if at all) in the context of QG.

[4] We believe the self-consistency story is relevant to the unification program, regardless of "the problem of quantum mechanics."

Using this restriction we compute the two-source transition amplitude over a graph with $N$ nodes and ($3N/2 - 2$) links, i.e., a "ladder" structure. The result is non-trivial, but the phase $\Phi$ contains three distinct parts: $\Phi_S$ involving only spatial links, $\Phi_T$ involving only temporal links and $\Phi_{ST}$ involving a complex mix of spatial and temporal links. In order to understand the empirical consequences of this solution, we analyze our solution using the simple twin-slit experiment. In this case, $\Phi_{ST} = 0$ and $\Phi_S + \Phi_T$ has a transparent form that suggests the square of a spatial link represents fundamental, relational units of length which thereby underwrites wave-particle duality.

As stated supra, one ontological implication of our approach to QM is that, fundamentally speaking, there are no mediating causal or dynamical entities such as particles or waves propagating from the source and impinging on the detector to "cause" detector clicks. Similarly, the implication for QFT is that fields are merely part of the computational device for producing $Z$, i.e., they are without fundamental ontic significance; instead, $\Sigma$ is the fundamental ontic structure representing the entire spatiotemporal configuration of the experiment (from preparation to outcomes). We begin in section 2 by motivating our methodology.

## 2. METHODOLIGICAL MOTIVATION

Our empirical goal is to tell a unified story about detector clicks—how they're distributed in space (e.g., interference patterns, interferometer outcomes, spin measurements), how they're distributed in time (e.g., click rates, coincidence counts), how they're distributed in space and time (e.g., particle trajectories), and how they generate more complex phenomena (e.g., photoelectric effect, superconductivity). Many in the foundations community are rightfully dubious that the unification program of particle physics will bear on the foundational problems of QM, given experiments such as those violating Bell's inequality. Thus, we start our program trying to explain a single detector click per RBW.

Specifically, we seek an approach whereby space, time and dynamical objects are relationally co-defined (arguments for this approach are in Stuckey *et al.*, 2008). In response to this challenge, we use the path integral formalism since it embodies relationalism in a necessary fashion, i.e., the computation of the transition amplitude $Z$ is based on the fact that "the source will emit and the detector receive" (Feynman, 1965, 167); per Tetrode, "the sun would not radiate if it were alone in space and no other bodies could absorb its radiation" (Tetrode, 1922, 325)[5]. Additionally, we employ the path integral approach over graphs since graph theory provides flexibility in the explicit co-construct of sources, space and time (as we will show), and it has already been shown to provide an excellent mathematics for the construct of a discrete basis to quantum physics (e.g., Markopoulou & Smolin, 2004).

Formally, it is not difficult to understand the difference in our proposed approach from that of conventional QM, essentially we see QM as spatially discrete QFT in the following sense. In the conventional path integral formalism for QM one starts with the amplitude for the propagation from the initial configuration space point $q_I$ to the final configuration space point $q_F$ in time $T$ via the unitary operator $e^{-iHT}$ (Zee, 2006), i.e., $\langle q_F | e^{-iHT} | q_I \rangle$. Breaking the time $T$ into $N$ pieces $\delta t$ and inserting the identity between each pair of operators $e^{-iH\delta t}$ via the complete set $\int dq|q\rangle\langle q| = 1$ we have

$$\langle q_F | e^{-iHT} | q_I \rangle = \left[ \prod_{j=1}^{N-1} \int dq_j \right] \langle q_F | e^{-iH\delta t} | q_{N-1} \rangle \langle q_{N-1} | e^{-iH\delta t} | q_{N-2} \rangle \ldots \langle q_2 | e^{-iH\delta t} | q_1 \rangle \langle q_1 | e^{-iH\delta t} | q_I \rangle$$

With $H = \frac{\hat{p}^2}{2m} + V(\hat{q})$ and $\delta t \rightarrow 0$ one can then show that the amplitude is given by

$$\langle q_F | e^{-iHT} | q_I \rangle = \int Dq(t) \exp\left[ i \int_0^T dt L(\dot{q}, q) \right] \tag{1}$$

where $L(\dot{q}, q) = \frac{1}{2} m\dot{q}^2 - V(q)$. When $q$ is the spatial coordinate on a detector transverse to the line joining source and detector, then $\prod_{j=1}^{N-1}$ can be thought of as $N$-1 'intermediate' detector surfaces interposed between the source and the final (real) detector, and $\int dq_j$ can be thought of all possible detection sites on the $j^{\text{th}}$ intermediate detector surface. In the continuum limit, these become $\int Dq(t)$ which is therefore viewed as a "sum over all possible paths" from the source to a particular point on the (real) detector, thus the term "path integral formalism."

[5] The path integral formalism requires both an emission event and a reception event; the formalism was motivated by the idea of treating advanced and retarded potentials equally.

Conversely, one obtains QFT by associating $q$ with the oscillator displacement at a *particular point* in space ($V(q) = \mathrm{k}q^2/2$) and taking the limit $\delta x \rightarrow 0$ so that space is filled with oscillators. Adding the spatial continuity is accounted for mathematically via $q_i(t) \rightarrow q(t,x)$, which is denoted $\varphi(t,x)$ and called a "field." Our amplitude now looks like

$$Z = \int D\varphi \exp\left[i\int d^4xL(\dot{\varphi},\varphi)\right] \tag{2}$$

where $L(\dot{\varphi},\varphi) = \frac{1}{2}(d\varphi)^2 - V(\varphi)$. Impulses $J$ are located in the field to account for particle creation and annihilation; these $J$ are called "sources" and we have $L(\dot{\varphi},\varphi) = \frac{1}{2}(d\varphi)^2 - V(\varphi) + J(t,x)\varphi(t,x)$. Rewriting this as $L(\dot{\varphi},\varphi) = \frac{1}{2}\varphi D\varphi + J(t,x)\varphi(t,x)$, where $D$ is a differential operator, and returning to a discrete spacetime lattice (typically, but not necessarily, hypercubical), $D \rightarrow \vec{\vec{A}}$, $J(t,x) \rightarrow \vec{J}$ (each component of which is associated with a point on the spacetime lattice) and $\varphi \rightarrow \vec{Q}$ (each component of which is associated with a point on the spacetime lattice). The discrete counterpart to Eq. (2) is then (Zee, 2003, 18)

$$Z = \int \ldots \int dQ_1 \ldots dQ_N \exp\left[\frac{i}{2}\vec{Q}\cdot\vec{\vec{A}}\cdot\vec{Q} + i\vec{J}\cdot\vec{Q}\right] \tag{3}.$$

In conventional quantum physics, QM is understood as (0+1)-dimensional QFT. We agree with that characterization but point out that it is at conceptual odds with our derivation of Eq. (1) when $\int Dq(t)$ represented a sum over all paths in space, i.e., when $q$ was understood as a location in space (specifically, a location along a detector surface). If QM is (0+1)-dimensional QFT, then $q$ is a *field displacement at a single location in space*. In that case, $\int Dq(t)$ must represent a sum over all field values at a particular point on the detector, not a sum over all paths through space from the source to a particular point on the detector. So, how *do* we relate a point on the detector to the source?

In answering this question, we now explain the difference between conventional QM and our proposed approach, highlighting its acausal, blockworld nature. Roughly, one might say we're using a modified QFT to link discrete sources $\vec{J}$ with one part of $\vec{J}$ used for the QM source and the other part of $\vec{J}$ used for the detector click; instead of $\delta x$ → 0, as in QFT, we assume $\delta x$ is measureable for QM phenomena. More specifically, we propose starting with Eq. (3) (modified to include scattering, spin, etc. where necessary) whence QM obtains in the limit $\delta t$ → 0, as in deriving Eq. (1), and QFT obtains in the additional limit $\delta x$ → 0, as in deriving Eq. (2). The QFT limit is well understood as it is the basis for lattice gauge theory, so one might argue that we're simply *clarifying* the QM limit where the path integral formalism is not widely employed. We counter such an argument by pointing out that Eq. (3) is *fundamental* in our approach, so *Eq. (3) is not a discrete approximation of Eqs. (1) & (2)*, but rather *Eqs. (1) & (2) are continuous approximations of Eq. (3)*. Placing the discrete formalism at bottom introduces conceptual and analytical differences. Conceptually, Eq. (1) of QM represents a sum over all field values at a particular point on the detector, while Eq. (3) of RBW is a 'mathematical machine' that measures the symmetry (strength of stationary points) contained in the kernel of the discrete action

$$\Sigma = \frac{1}{2}\bar{\bar{A}} + \vec{J} \tag{4}$$

This kernel or *actional* yields the discrete action after operating on a particular vector $\vec{Q}$ (field). The actional represents a *fundamental, spatiotemporally holistic description of the experimental arrangement, to include outcomes,* and $Z$ is a measure of its symmetry. For this reason, we prefer to call $Z$ the *symmetry amplitude* of the spatiotemporal experimental configuration. Notice that in this view, fields have no ontic significance—they are merely part of the computational device for measuring the symmetry of Σ (representing what *is* ontically significant at the fundamental level). Analytically, because we are *starting* with a discrete formalism, we are in position to mathematically explicate trans-temporal identity, whereas this process is unarticulated elsewhere in physics (as elaborated immediately below). As we will now see, this leads to our proposed self-consistency criterion underlying Eq. (3).

The QM limit $\delta t \rightarrow 0$ of Eq. (3) results in a spatially discrete distribution of interacting sources $J_i(t)$ and illustrates a key aspect of RBW ontology, i.e., interaction without mediation—there is an interaction of sources without mediating waves or particles traveling through intervening space (notice also that there is no field between sources). The spatiotemporally discrete formalism also illustrates nicely how QM tacitly assumes an *a priori* process of trans-temporal identification, $\vec{J} \rightarrow J_i(t)$. Indeed, there is no principle which dictates the construct of diachronic entities fundamental to the formalism of dynamics in general—these objects are "put in by hand" throughout physics. When Albrecht and Iglesias (2008) allowed time to be an "internal variable" after quantization, as in the Wheeler-DeWitt equation, they found "there is no one set of laws, but a whole library of different cosmic law books" (Siegfried, 2008). They called this the "clock ambiguity." In order to circumvent this "arbitrariness in the predictions of the theory" they proposed that "the principle behind the regularities that govern the interaction of entities is … the idea that individual entities exist at all" (Siegfried, 2008). Albrecht and Iglesias characterize this as "the central role of quasiseparability." Similarly, the RBW approach requires a fundamental principle whence the trans-temporal identity employed tacitly in QM and all dynamical theories. Our graphical starting point does not contain dynamical entities, space or time *per se* so we must formalize counterparts to these concepts. Clearly, the process $\vec{J} \rightarrow J_i(t)$ is an organization of the set $\vec{J}$ on two levels—there is the split of the set into *i* subsets, one for each source, and there is the ordering *t* over each subset. The split represents space, the ordering represents time and the result is objecthood. In this sense, space, time and 'things' (trans-temporal objects "made of" matter) are co-defined in our formalism. Consequently, we believe the articulation of the otherwise tacit construct of dynamical entities has a mathematical counterpart fundamental to the action, which is in accord with Toffoli's belief that there exists a mathematical tautology fundamental to the action (Toffoli, 2003):

> Rather, the motivation is that principles of great generality must be by their very nature *trivial*, that is, expressions of broad tautological identities. If the principle of least action, which is so general, still looks somewhat mysterious, that means we still do not understand what it is really an expression of—what it is trying to tell us.

Our use of a self-consistency criterion (SCC) is not without precedence, as we already have an ideal example in Einstein's equations of general relativity (GR). Momentum, force and energy all depend on spatiotemporal measurements (tacit or explicit), so the stress-energy tensor cannot be constructed without tacit or explicit knowledge of the spacetime metric (technically, the stress-energy tensor can be written as the functional derivative of the matter-energy Lagrangian with respect to the metric). But, if one wants a 'dynamic' spacetime in the parlance of GR, the spacetime metric must depend on the matter-energy distribution in spacetime. GR solves this dilemma by demanding the stress-energy tensor be 'consistent' with the spacetime metric per Einstein's equations. This self-consistency hinges on divergence-free sources, which finds a mathematical counterpart in the topological maxim, "the boundary of a boundary is zero" (Misner *et al.*, 1973). So, Einstein's equations of GR provide an example of an SCC. In fact, our SCC is based on the same topological maxim for the same reason, as are quantum and classical electromagnetism (Misner *et al.*, 1973; Wise, 2006).

In order to explore the mathematical co-definition of space, time and 'things', we will use graph theory *a la* Wise (2006) and show that $\partial_1\partial_1^T$ , where $\partial_1$ is a boundary operator in the chain complex of our graph satisfying $\partial_1\partial_2 = 0$, has precisely the same form as the matrix operator in the discrete action for coupled harmonic oscillators. Therefore, we are led to speculate that $\vec{\vec{A}} \propto \partial_1\partial_1^T$. Defining the source vector $\vec{J}$ relationally via links of the graph per $\vec{J} \propto \partial_1\vec{e}$ then gives tautologically $\vec{\vec{A}}\vec{v} \propto \vec{J}$, where $\vec{e}$ is the vector of links and $\vec{v}$ is the vector of vertices. $\vec{\vec{A}}\vec{v} \propto \vec{J}$ constitutes what is meant by a self-consistent co-definition of space, time and dynamical objects and thereby constrains $\vec{\vec{A}}$ and $\vec{J}$ in the actional. Thus, our self-consistency criterion $\vec{\vec{A}}\vec{v} \propto \vec{J}$ provides a basis for the discrete action in accord with Toffoli and supports our view that Eq. (3) is fundamental to Eqs. (1) & (2), rather than the converse and, conceptually, that is the basis of our proposed formalism.

**3. THE PROPOSED FORMALISM**

*3.1 The Two-Source Symmetry Amplitude*. In order to motivate our general results, we will first consider a simple graph with six vertices, seven links and two plaquettes (cells) for our 2D spacetime model (Figure 1). Our goal with this simple model is to seek relevant structure that might be used to infer a self-consistency criterion. We begin by constructing the boundary operators over our graph.

The boundary of $\mathbf{p}_1$ is $\mathbf{e}_4 + \mathbf{e}_5 - \mathbf{e}_2 - \mathbf{e}_1$, which also provides an orientation. The boundary of $\mathbf{e}_1$ is $\mathbf{v}_2 - \mathbf{v}_1$, which likewise provides an orientation. Using these conventions for the orientations of links and plaquettes we have the following boundary operator for $C_2 \rightarrow C_1$, i.e., space of plaquettes mapped to space of links in the spacetime chain complex:

$$\partial_2 = \begin{bmatrix} -1 & 0 \\ -1 & 1 \\ 0 & -1 \\ 1 & 0 \\ 1 & 0 \\ 0 & 1 \\ 0 & -1 \end{bmatrix} \tag{5}$$

Notice the first column is simply the links for the boundary of $\mathbf{p}_1$ and the second column is simply the links for the boundary of $\mathbf{p}_2$. We have the following boundary operator for $C_1 \rightarrow C_0$, i.e., space of links mapped to space of vertices in the spacetime chain complex:

$$\partial_1 = \begin{bmatrix} -1 & 0 & 0 & -1 & 0 & 0 & 0 \\ 1 & -1 & -1 & 0 & 0 & 0 & 0 \\ 0 & 0 & 1 & 0 & 0 & 0 & -1 \\ 0 & 0 & 0 & 1 & -1 & 0 & 0 \\ 0 & 1 & 0 & 0 & 1 & -1 & 0 \\ 0 & 0 & 0 & 0 & 0 & 1 & 1 \end{bmatrix} \tag{6}$$

which completes the spacetime chain complex, $C_0 \xleftarrow{\partial_1} C_1 \xleftarrow{\partial_2} C_2$. Notice the columns are simply the vertices for the boundaries of the edges. These boundary operators satisfy $\partial_1\partial_2 = 0$ as required for "boundary of a boundary is zero." We want our SCC ultimately founded on this topological maxim so we construct our actional from the boundary operators of our spacetime chain complex. The manner by which we do this is suggested by the discrete action for coupled harmonic oscillators on our simple graph.

The potential for coupled oscillators can be written

$$V(q_1,q_2) = \sum_{a,b}\frac{1}{2}k_{ab}q_a q_b = \frac{1}{2}kq_1^2 + \frac{1}{2}kq_2^2 + k_{12}q_1q_2 \tag{7}$$

where $k_{11} = k_{22} = k$ (positive) and $k_{12} = k_{21}$ (negative) per the classical analogue (Figure 2) with $k = k_1 + k_3 = k_2 + k_3$ and $k_{12} = -k_3$ to recover the form in Eq. (7). The Lagrangian is then

$$L = \frac{1}{2}m\dot{q}_1^2 + \frac{1}{2}m\dot{q}_2^2 - \frac{1}{2}kq_1^2 - \frac{1}{2}kq_2^2 - k_{12}q_1q_2 \tag{8}$$

so our QM symmetry amplitude is

$$Z = \int Dq(t)\exp\left[-\int_0^T dt\left[\frac{1}{2}m\dot{q}_1^2 + \frac{1}{2}m\dot{q}_2^2 + V(q_1,q_2) - J_1q_1 - J_2q_2\right]\right] \tag{9}$$

after Wick rotation. This gives

$$\vec{\vec{A}} = \begin{bmatrix} \left(\frac{m}{\Delta t}+k\Delta t\right) & \frac{-m}{\Delta t} & 0 & k_{12}\Delta t & 0 & 0 \\ \frac{-m}{\Delta t} & \left(\frac{2m}{\Delta t}+k\Delta t\right) & \frac{-m}{\Delta t} & 0 & k_{12}\Delta t & 0 \\ 0 & \frac{-m}{\Delta t} & \left(\frac{m}{\Delta t}+k\Delta t\right) & 0 & 0 & k_{12}\Delta t \\ k_{12}\Delta t & 0 & 0 & \left(\frac{m}{\Delta t}+k\Delta t\right) & \frac{-m}{\Delta t} & 0 \\ 0 & k_{12}\Delta t & 0 & \frac{-m}{\Delta t} & \left(\frac{2m}{\Delta t}+k\Delta t\right) & \frac{-m}{\Delta t} \\ 0 & 0 & k_{12}\Delta t & 0 & \frac{-m}{\Delta t} & \left(\frac{m}{\Delta t}+k\Delta t\right) \end{bmatrix} \tag{10}$$

on our graph. Thus, we borrow (loosely) from Wise (2006) and suggest $\vec{\vec{A}} \propto \partial_1\partial_1^T$ since

$$\partial_1\partial_1^T = \begin{bmatrix} 2 & -1 & 0 & -1 & 0 & 0 \\ -1 & 3 & -1 & 0 & -1 & 0 \\ 0 & -1 & 2 & 0 & 0 & -1 \\ -1 & 0 & 0 & 2 & -1 & 0 \\ 0 & -1 & 0 & -1 & 3 & -1 \\ 0 & 0 & -1 & 0 & -1 & 2 \end{bmatrix} \qquad (11)$$

produces precisely the same form as Eq. (10) and quantum theory is known to be "rooted in this harmonic paradigm" (Zee, 2003, 5). [In fact, these matrices will continue to have the same form as one increases the number of vertices in Figure 1.] Now we construct a suitable candidate for $\vec{J}$, relate it to $\vec{\vec{A}}$ and infer our SCC.

Recall that $\vec{J}$ has a component associated with each node so here it has components, $J_n$, n = 1, 2, …, 6; $J_n$ for n = 1, 2, 3 represents one source and $J_n$ for n = 4, 5, 6 represents the second source. We propose $\vec{J} \propto \partial_1\vec{e}$ ,where $e_i$ are the links of our graph, since

$$\partial_1\vec{e} = \begin{bmatrix} -1 & 0 & 0 & -1 & 0 & 0 & 0 \\ 1 & -1 & -1 & 0 & 0 & 0 & 0 \\ 0 & 0 & 1 & 0 & 0 & 0 & -1 \\ 0 & 0 & 0 & 1 & -1 & 0 & 0 \\ 0 & 1 & 0 & 0 & 1 & -1 & 0 \\ 0 & 0 & 0 & 0 & 0 & 1 & 1 \end{bmatrix} \begin{bmatrix} e_1 \\ e_2 \\ e_3 \\ e_4 \\ e_5 \\ e_6 \\ e_7 \end{bmatrix} = \begin{bmatrix} -e_1 - e_4 \\ e_1 - e_2 - e_3 \\ e_3 - e_7 \\ e_4 - e_5 \\ e_2 + e_5 - e_6 \\ e_6 + e_7 \end{bmatrix} \qquad (12)$$

provides a means of understanding vertices in terms of links and ultimately we want sources defined relationally. For example, vertex 1 is the origin of both links 1 and 4, and the first entry of $\partial_1\vec{e}$ is $-e_1 - e_4$ (negative/positive means the link starts/ends at that vertex). Since $J_n$ are associated with the vertices to represent 'things', $\vec{J} \propto \partial_1\vec{e}$ is a graphical representation of "relata from relations." [Note: $\partial_1\vec{e}$, which we denote $\vec{v}\,*$ and associate with $\vec{v}$, is not equal to $\vec{v}$ proper.]

With these definitions of $\vec{\vec{A}}$ and $\vec{J}$ we have, *ipso facto*, $\vec{\vec{A}}\vec{v} \propto \vec{J}$ as the basis of our SCC since

$$\partial_1\partial_1^T\vec{v}=\begin{bmatrix}2&-1&0&-1&0&0\\-1&3&-1&0&-1&0\\0&-1&2&0&0&-1\\-1&0&0&2&-1&0\\0&-1&0&-1&3&-1\\0&0&-1&0&-1&2\end{bmatrix}\begin{bmatrix}v_1\\v_2\\v_3\\v_4\\v_5\\v_6\end{bmatrix}=\begin{bmatrix}2v_1-v_2-v_4\\-v_1+3v_2-v_3-v_5\\-v_2+2v_3-v_6\\-v_1+2v_4-v_5\\-v_2-v_4+3v_5-v_6\\-v_3-v_5+2v_6\end{bmatrix}=\begin{bmatrix}-e_1-e_4\\e_1-e_2-e_3\\e_3-e_7\\e_4-e_5\\e_2+e_5-e_6\\e_6+e_7\end{bmatrix}=\partial_1\vec{e}=\vec{v}^* \quad (13)$$

where we've used $e_1 = v_2 - v_1$ (etc.) to obtain the last column, which constitutes a definition of links in terms of vertices. Thus, the SCC $\vec{\vec{A}}\vec{v} \propto \vec{J}$ obtains tautologically via the maxim, "the boundary of a boundary is zero," as desired.

Moving now to $N$ dimensions, the solution of Eq. (3) is

$$Z=\left(\frac{(2i\pi)^N}{\det(A)}\right)^{1/2}\exp\left[-\frac{i}{2}\vec{J}\cdot\vec{\vec{A}}^{-1}\cdot\vec{J}\right] \quad (14).$$

Using $\vec{J}=\alpha\partial_1\vec{e}$ and $\vec{\vec{A}}=\beta\partial_1\partial_1^T$ ($\alpha, \beta \in \mathbb{R}$) with the SCC gives $\vec{\vec{A}}\vec{v}=\frac{\beta}{\alpha}\vec{J}$, so that $\vec{v}=\frac{\beta}{\alpha}\vec{\vec{A}}^{-1}\vec{J}$. However, $\vec{\vec{A}}^{-1}$ doesn't exist because $\vec{\vec{A}}$ is singular, which means of course that Eq. (3) is ill-defined for this problem. $\vec{\vec{A}}$ is singular because one of its eigenvalues is zero, therefore the row space of $\vec{\vec{A}}$ is an ($N$-1)-dimensional hyperplane of the $N$-dimensional vector space. [The eigenvector with eigenvalue of zero, i.e., normal to this hyperplane, is $[1,1,1,\ldots,1]^T$.] Since $\vec{J}$ resides in this ($N$-1)-dimensional hyperplane as well (which you can see from $\sum_i J_i = 0$), we propose restricting the path integral of Eq. (3) to the row space of $\vec{\vec{A}}$, i.e.,

$$Z=\int_{-\infty}^{\infty}\ldots\int_{-\infty}^{\infty}d\tilde{Q}_1\ldots d\tilde{Q}_{N-1}\exp\left[\sum_{j=1}^{N-1}\left(\frac{i}{2}\tilde{Q}_j^2a_j+i\tilde{J}_j\tilde{Q}_j\right)\right] \quad (15)$$

where $\tilde{Q}_j$ are the coordinates associated with the eigenbasis of $\vec{\vec{A}}$ and $\tilde{Q}_N$ is associated with eigenvalue zero, $a_j$ is the eigenvalue of $\vec{\vec{A}}$ corresponding to $\tilde{Q}_j$, and $\tilde{J}_j$ are the components of $\vec{J}$ in the eigenbasis of $\vec{\vec{A}}$. Again, on our view, $Z$ does not reflect a "sum over all paths in configuration space," but rather it is a 'mathematical machine' which produces a relative symmetry amplitude for the various Σ associated with different experimental outcomes and configurations. Thus, our path integral restriction is supported conceptually as well as formally and revises Eq. (14) to read

$$Z = \left( \frac{(2i\pi)^{N-1}}{\prod_{j=1}^{N-1} a_j} \right)^{1/2} \prod_{j=1}^{N-1} \exp\left[ -\frac{i}{2} \frac{\tilde{J}_j^2}{a_j} \right] \tag{16}$$

Since $\vec{J}$ is defined via links we have characterized the symmetry amplitude in terms of relations and the non-zero eigenvalues of $\vec{\vec{A}}$.

As an aside, we note that Eq. (16) also obtains in the classical limit, $\hbar \rightarrow 0$, of Eq. (15) via the stationary phase method (Zee, 2006, 15). That is,

$$\int_{-\infty}^{\infty} \ldots \int_{-\infty}^{\infty} d\tilde{Q}_1 \ldots d\tilde{Q}_{N-1} \exp\left[ \frac{i}{\hbar} f(\vec{\tilde{Q}}) \right] = \exp\left[ \frac{i}{\hbar} f(\vec{Q}_E) \right] \left( \frac{(2\pi\hbar i)^{N-1}}{\det(f''(\vec{Q}_E))} \right)^{1/2} \tag{17}$$

where $\vec{Q}_E$ is the extremum of $f\left(\vec{\tilde{Q}}\right)$, which is at most quadratic in $\vec{\tilde{Q}}$. We have

$$f\left(\vec{\tilde{Q}}\right) = \sum_{j=1}^{N-1} \left( \frac{1}{2} \tilde{Q}_j^2 a_j + \tilde{J}_j \tilde{Q}_j \right) \tag{18}$$

which has an extremum at $\tilde{Q}_i = \dfrac{-\tilde{J}_i}{a_i}$ so

$$f\left(\vec{Q}_E\right) = \sum_{j=1}^{N-1} \left( \frac{-\tilde{J}_j^2}{2a_j} \right) \tag{19}$$

Since $\dfrac{\partial^2 f}{\partial \tilde{Q}_i^2} = a_i$, Eqs. (19) & (17) give Eq. (16) with $\hbar$ restored. Thus our spatially discrete QFT version of QM corresponds to the standard path integral formulation of QM where the potential has the form $V = a + bx + cx^2 + e\dot{x} + gx\dot{x}$ (Shankar, 1994, 231). In fact, we chose $\vec{\vec{A}} = \beta \partial_1 \partial_1^T$ precisely because it reproduces the action for coupled harmonic oscillators and therein *V* is quadratic in *q*. However, keep in mind that *q* is *not* the spatial location *x* of a particle in the potential *V* as is standard in QM, but *q* is the *field value at a point in space* as is standard in QFT. And, at our proposed fundamental level, it is Σ that provides the basic ontological depiction of the experiment and $q$ is merely part of the mathematical machinery used to evaluate Σ.

Returning to Eq. (16), we find in general that half the eigenvectors of $\vec{\vec{A}}$ are of the form $\begin{bmatrix} \vec{x} \\ \vec{x} \end{bmatrix}$ and half are of the form $\begin{bmatrix} \vec{x} \\ -\vec{x} \end{bmatrix}$. The eigenvalues are given by $\lambda \pm 1$ where $\lambda - 1$ is the eigenvalue for $\begin{bmatrix} \vec{x} \\ \vec{x} \end{bmatrix}$, $\lambda + 1$ is the eigenvalue for $\begin{bmatrix} \vec{x} \\ -\vec{x} \end{bmatrix}$, and

$\lambda_j = 3 - 2\cos\left(\frac{j2\pi}{N}\right),\ \ j = 0,\ldots,\left(\frac{N}{2} - 1\right)$. The $k$ components of $\vec{x}$ for a given $\lambda_j$ are

$x_{jk} = \sqrt{\frac{2}{N}}\cos\left(\frac{j(2k-1)\pi}{N}\right),\ \ k = 1,\ldots,\frac{N}{2}$ for $j > 0$ and $x_{0k} = \sqrt{\frac{1}{N}},\ \ k = 1,\ldots,\frac{N}{2}$ for $j = 0$

($j = 0 \rightarrow$ eigenvalues of $\vec{\vec{A}}$ are 0 and 2). We have $N$ nodes and ($3N/2 - 2$) links. Define the temporal (vertical) links $e_i$ in terms of vertices $v_i$ in the following fashion:

$$e_i = v_{i+1} - v_i \qquad \text{i} = 1 \text{ to } N/2 - 1$$

and

$$e_{\frac{N}{2}+i-1} = v_{\frac{N}{2}+i+1} - v_{\frac{N}{2}+i} \qquad \text{i} = 1 \text{ to } N/2 - 1.$$

Define the spatial (horizontal) links via:

$$e_{N+i-2} = v_{\frac{N}{2}+i} - v_i \quad \text{i} = 1 \text{ to } N/2.$$

This gives

$$\vec{J} = \begin{bmatrix} -e_1 - e_{N-1} & \\ -e_i + e_{i-1} - e_{N+i-2} & i = 2,\ldots\frac{N}{2} - 1 \\ e_{\frac{N}{2}-1} - e_{N+\frac{N}{2}-2} & \\ e_{N-1} - e_{\frac{N}{2}} & \\ e_{\frac{N}{2}+i-2} + e_{N+i-2} - e_{\frac{N}{2}+i-1} & i = 2,\ldots\frac{N}{2} - 1 \\ e_{N+\frac{N}{2}-2} + e_{N-2} & \end{bmatrix} \qquad (20).$$

We then need to find the projection of $\vec{J}$ on each of the orthonormal eigenvectors of $\vec{\vec{A}}$ that have non-zero eigenvalues. Call each projection $\tilde{J}_i = \langle i|J\rangle$, where $\langle i|$ is the i^th^ orthonormal eigenvector. Let $a_i$ ($i$ = 1, $N$-1) be the non-zero eigenvalues of $\vec{\vec{A}}$ associated with the eigenvectors $\langle i|$, ($i$ = 1, $N$-1), respectively. To complete the two-source symmetry amplitude we need to compute the phase

$$\Phi = -\sum_{i=1}^{N-1} \frac{\left(\tilde{J}_i\right)^2}{2a_i\hbar\beta} \tag{21}$$

where $\hbar$ is viewed as a fundamental scaling factor with the dimensions of action. We find $\Phi = -(\Phi_S + \Phi_T + \Phi_{ST})/(2\hbar\beta)$, where

$$\Phi_S = \frac{2\alpha^2}{N}\left[\sum_{k=1}^{\frac{N}{2}-1} e_{k+N-2}\right]^2 \tag{22}$$

involves only spatial links

$$\Phi_T = \sum_{j=1}^{\frac{N}{2}-1} \frac{2\alpha^2}{N}\left[\sum_{k=1}^{\frac{N}{2}-1}\left(e_k + e_{k+\frac{N}{2}-1}\right)\sin\left(\frac{jk2\pi}{N}\right)\right]^2 \tag{23}$$

involves only temporal links and

$$\Phi_{ST} = \sum_{j=1}^{\frac{N}{2}-1} \frac{4\alpha^2}{N\left(1+\sin^2\frac{j\pi}{N}\right)}\left[\sin\frac{j\pi}{N}\sum_{k=1}^{\frac{N}{2}-1}\left(e_k - e_{k+\frac{N}{2}-1}\right)\sin\left(\frac{jk2\pi}{N}\right) + \sum_{k=1}^{\frac{N}{2}}\left(e_{k+N-1}\right)\cos\left(\frac{(2k-1)j\pi}{N}\right)\right]^2 \tag{24}$$

involves mix of spatial and temporal links.

In summary: SCC ($\vec{\vec{A}}\vec{v} \propto \vec{J}$) → actional ($\Sigma = \frac{1}{2}\vec{\vec{A}} + \vec{J}$) → symmetry amplitude ($Z$) → relative probability for a particular spatiotemporal configuration and outcome when normalized over all possible configurations and outcomes of interest. This initiates the development of an analytical and foundational basis for the RBW ontology and methodology, i.e., a discrete graph theoretic approach to quantum physics, thereby rendering a first payment on the promissory note.

*3.2 The Twin-Slit Experiment.* The simple twin-slit experiment is used for a preliminary study of our two-source amplitude. We point out that the potential $V$ is zero in QM for this case (free-particle propagator) while our discrete QFT Eq. (15) is quadratic in the field so, again, one should not confuse $q$ with the position of a particle in space. We begin with what we already know of this idealized situation per QM, then we make inferences concerning our graph structure via Eqs. (22) – (24).

For a free particle of mass $m$ we have from QM (Shankar, 1994)

$$\psi = A\sqrt{\frac{m}{2\pi\hbar it}}\exp\left[\frac{imx^2}{2\hbar t}\right] \propto \exp\left[\frac{imx^2 t}{2\hbar t^2}\right] = \exp\left[\frac{imv^2 t}{2\hbar}\right] = \exp\left[\frac{ipvt}{2\hbar}\right] = \exp\left[\frac{iv_\varphi t}{\lambda}2\pi\right] \quad (25)$$

where $v_\varphi$ is the phase velocity and equal to half the particle velocity (Park, 1992) and $\psi(\mathrm{x},0) = A\delta(\mathrm{x} = 0)$. [The standard QM path integral produces a propagator and Eq. (25) is obtained from it by connecting a point source to a point at the detector, each of these points is understood to be half of our source vector $\vec{J}$, thus our use of the two-source symmetry amplitude.] Using Eq. (25), the twin-slit interference pattern is given by

$$|\psi_1 + \psi_2|^2 \propto \left|\exp\left[\frac{iv_\varphi t_1}{\lambda}2\pi\right] + \exp\left[\frac{iv_\varphi t_2}{\lambda}2\pi\right]\right|^2 = 2 + 2\cos\left[\frac{v_\varphi(t_1 - t_2)}{\lambda}2\pi\right] \quad (26)$$

and therefore maxima occur at angles where

$$v_\varphi(t_1 - t_2) = n\lambda \quad n \in \mathbb{Z} \quad (27).$$

For photons

$$\psi \propto \exp\left[\frac{iE(t_1 - t_2)}{\hbar}\right] = \exp\left[\frac{ihf(t_1 - t_2)}{\hbar}\right] = \exp\left[\frac{ic(t_1 - t_2)}{\lambda}2\pi\right] \quad (28)$$

so maxima occur at angles where

$$c(t_1 - t_2) = n\lambda \quad n \in \mathbb{Z} \quad (29).$$

Since a photon yields a single click (not a series of clicks whence a trajectory), $c$ cannot be directly measured for a photon just as $v_\varphi$ cannot be directly measured for a massive particle, so Eqs. (25) & (28) do not differ structurally. Since the experimental outcome (interference pattern) is time-independent and does not involve clicks linked temporally (explicit trajectories), QM's theoretical description of the interference pattern is purely kinematical (involves concepts of length, time and velocity, but not mass, momentum, force, energy, etc.). In order that Eqs. (22) – (24) make correspondence with QM in this case, we must have

$$\Phi = \left[\frac{vt}{\lambda}2\pi\right] \qquad (30)$$

where $v$ is simply a scaling factor between space and time in this purely geometric result. In the twin-slit experiment this means

$$\Phi_1 - \Phi_2 = \frac{v(t_1 - t_2)}{\lambda} \qquad (31).$$

Let $e_i$ be the links of graph 1 (whence $\Phi_1$) and $\tilde{e}_i$ the links of graph 2 (whence $\Phi_2$). We expect the temporal links of the source representing the click to be equal between graphs since these sources in both graphs represent one and the same click. We also expect the temporal links of the sources representing each slit to be equal since these sources are presumed coherent in the twin-slit experiment. Suppose further that all temporal links of either graph are equal to one another (nothing intrinsic to the experimental configuration requires variable clock rates), so we have $\tilde{e}_i = e_i = e_T$ for $i = 1$ to $N-2$, i.e., for the temporal links. We do expect the spatial links to differ between graphs, reflecting the different distances from each slit to a particular click location. Let us assume all spatial links of each graph are equal to one another (static situation) so we have $e_{N+i-2} = e_x$ and $\tilde{e}_{N+i-2} = \tilde{e}_x$ for $i = 1$ to $N/2$, i.e., for the spatial links. In this highly simplified case, we find $\Phi_{ST} = 0$ and

$$\Phi_S + \Phi_T = \frac{N}{2}\alpha^2 e_x^2 + (N-2)\alpha^2 e_T^2 \qquad (32).$$

Since $N/2$ is the number of spatial links and $N-2$ is the number of temporal links, this is the result that would've obtained if $\vec{\vec{A}}$ wasn't singular; $\vec{v} \sim \vec{\vec{A}}^{-1}\vec{J} \rightarrow \vec{J}\cdot\vec{\vec{A}}^{-1}\vec{J} \sim \vec{J}\cdot\vec{v}$ and $\vec{J}\cdot\vec{v} \sim \sum_{All} e_i^2$. We now have

$$\frac{v(t_1 - t_2)}{\lambda} = \frac{N\alpha^2}{4\hbar\beta}\left(e_x^2 - \tilde{e}_x^2\right) \tag{33}$$

(dropping the irrelevant negative sign). We can of course measure $\Delta\ell := v(t_1 - t_2)$ for any maximum of the twin-slit interference pattern and deduce $\lambda$, since in those cases $\lambda = n\Delta\ell$ per Eqs. (27) or (29).

Let us therefore suppose that $\lambda$ is the fundamental, relational unit of length for this particular pair of graphs. We have $[\alpha]$ = (momentum) and $[\beta]$ = (momentum)/(length), and Eqs. (27), (29) and (33) give us

$$\frac{N\alpha^2}{4\hbar\beta}\left(e_x^2 - \tilde{e}_x^2\right) = 2\pi n \qquad n \in \mathbb{Z} \tag{34}$$.

With $h$ the fundamental unit of action we infer $\alpha = h/\lambda$ and $\beta = h/\lambda^2$, so Eq. (34) gives us

$$\left(\frac{N}{2}\right)\frac{\left(e_x^2 - \tilde{e}_x^2\right)}{2} = n \qquad n \in \mathbb{Z} \tag{35}$$

for interference maxima. Eq. (35) implies $Ne_x^2/4$ can be thought of as the number of fundamental, relational length units (let us call them "waves") represented by the spatial part of the graph. In that case, since $N/2$ is the number of spatial links, $e_x^2/2$ is the number of waves represented by each spatial link.

While this analysis is highly heuristic given the underdetermination of variables at this point, it is a reasonable start and does suggest a formal basis for wave-particle duality, i.e., links are "waves" and collections thereof produce "particle" outcomes (clicks). Of course, Eqs. (22) – (24) are far more complex than the RHS of Eq. (30) and we are not suggesting they be used in place of Eqs. (26) – (29). Rather, in this context, we are leaning on the established result to provide analytical guidance for what we believe is the more fundamental approach; in return, the more fundamental formal result provides conceptual clarity to the established formal result. As we make progress analytically, we expect to move beyond providing conceptual clarity to already established formal results and bring our analytic technique to bear on unresolved formal issues, e.g., QG as mentioned in an earlier footnote.

**4. CONCLUSION**

We have presented the discrete, two-source transition amplitude for a ladder graph of arbitrary size. The computation required a path integral restriction to avoid a singularity, but we argued that the restriction is perfectly reasonable in our approach. This solution illustrates our proposed formalism fundamental to quantum mechanics (QM), whereby there are no mediating causal entities responsible for detector clicks. In a sense, we are viewing QM as a spatially discrete quantum field theory (QFT), although our spatiotemporally discrete formalism is fundamental to QFT as well as QM. In this approach, motivated by our interpretation of QM called *Relational Blockworld* (RBW), dynamical entities, space and time must be self-consistently co-defined. To codify this demand for self-consistency, we proposed a self-consistency criterion (SCC) in the context of discrete graph theory *a la* Wise (2006) that underlies the discrete action. The SCC constrains the kernel of the discrete action, and it is this kernel or "actional" which provides the fundamental, relational characterization of a particular experiment past, present and future to include outcomes. The transition (or symmetry) amplitude *Z* is then understood to measure the symmetry contained in the actional, rather than its typical interpretation as a sum over all field configurations. This subtle difference in the interpretation of Z justifies restricting our path integral to the row space of our discrete differential operator, which also contains our discrete source vector. The square of Z provides a relative probability for a particular experimental configuration and outcome when normalized over all possible configurations and outcomes of interest. Empirically speaking, the distribution of individual detector clicks is the fundamental observational fact we seek to explain. Thus, in our view, particle physics is in the business of characterizing spatiotemporal click patterns (trajectories), so trajectory characteristics such as mass, charge and spin are not to be reified as the properties of "click-causing particles" moving through the detector. Likewise, fields have no fundamental ontic status but are simply part of the computational machinery of *Z*. Our approach constitutes a *new basis* for QM as opposed to a mere *discrete approximation* thereto, since we are proposing a basis for the action (SCC), which is otherwise fundamental.

**Figure 1**

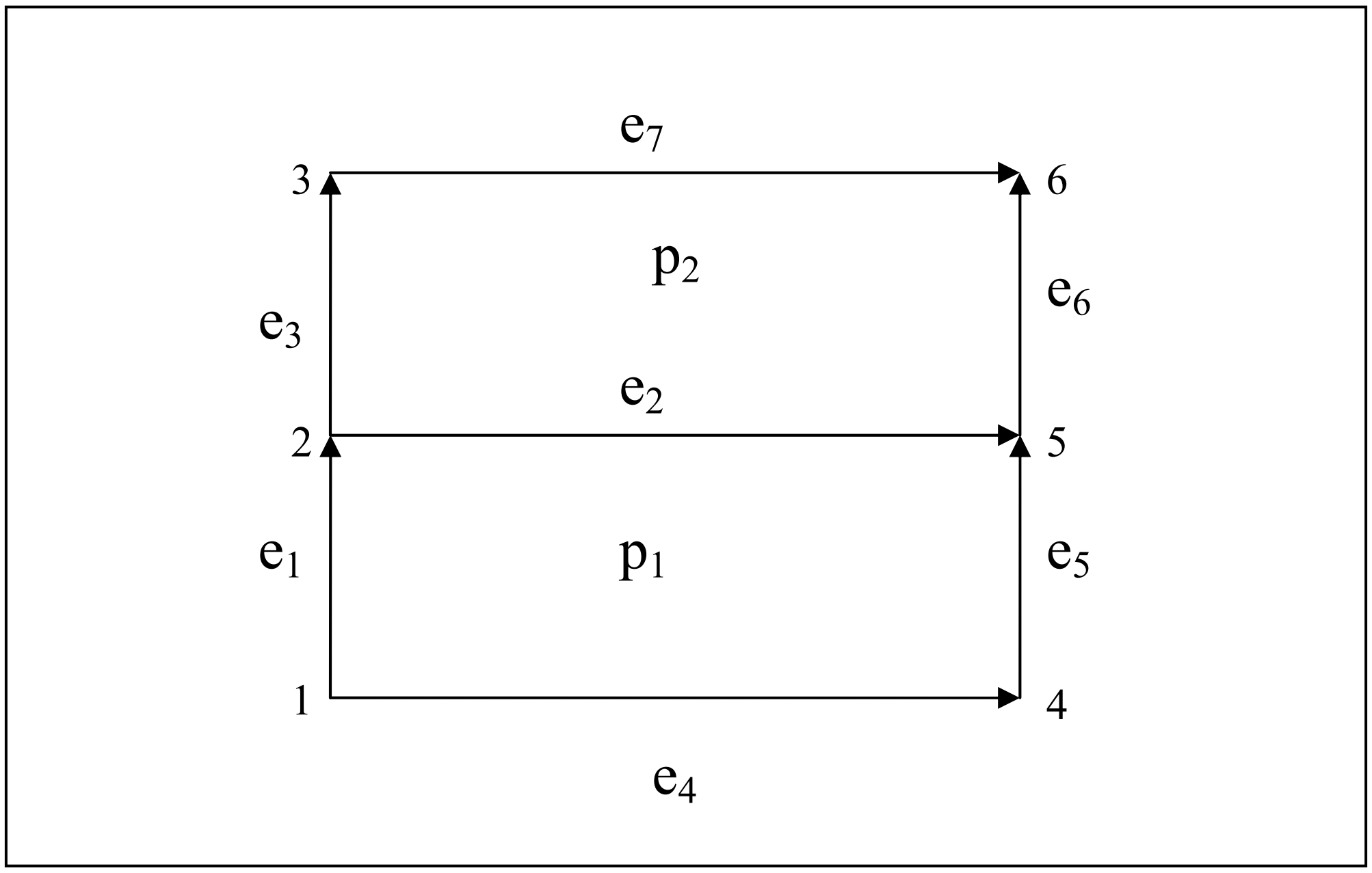

**Figure 2**

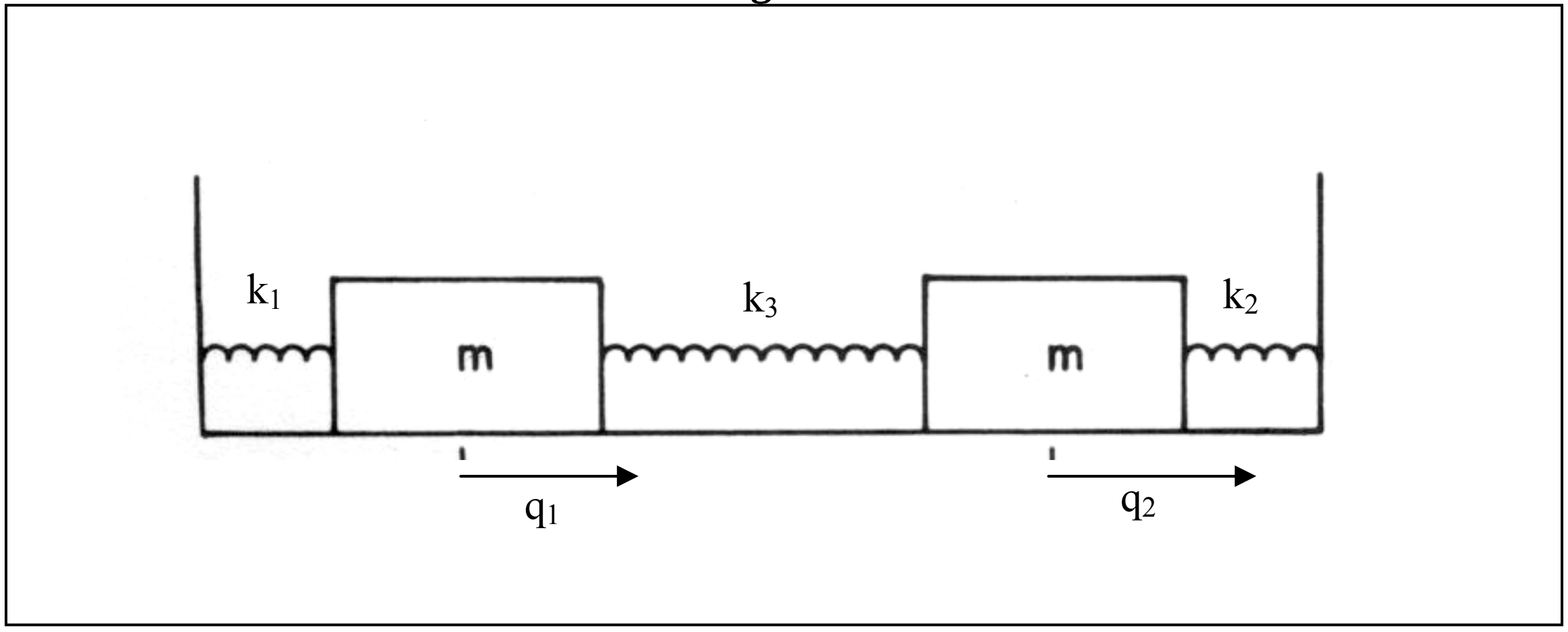